\documentstyle[12pt,titlepage,openbib]{article}
\begin{document}
\newcommand{\sg}{$\sigma$\ }
\newcommand{\ras}{$\rho_{\alpha,\sigma}$\ }
\newcommand{\rs}{$\rho_{\sigma}$\ }
\newcommand{\N}{$^{1,2}$}
\baselineskip 18pt

\section{Introduction}

The occurrence of close encounters between galaxies has recently been
recognized, by means of N-body simulations, as a possible mechanism for
triggering bars in spiral discs (Byrd et al., 1986, Noguchi, 1987, Gerin,
Combes, Athanassoula, 1990). However, so far little effort has been devoted to
the investigation of observational evidence for this effect. Thompson (1981)
noticed a greater fraction of barred spiral galaxies in the dense, inner
regions of the Coma cluster than in the outer regions; this has been
interpreted by Byrd and Valtonen (1990), always on the basis of N-body
simulations, as the result of the interaction with the strong tidal field of
the whole cluster. Noguchi (1987) recognized an excess of barred spirals among
the galaxies of the Arp Atlas (Arp, 1966), which are presumably interacting,
compared with the galaxies listed in the RC2 catalogue (de Vaucouleurs et al.
1976).

The only relevant extensive studies are due to Kumai et al. (1986) and
Elmegreen, Elmegreen, Bellin (1990; hereafter EEB). Relying on the
morphological  classifications reported in the UGC catalogue of galaxies
(Nilson, 1973), the Japanese authors counted the number of the UGC barred and
unbarred spiral galaxies located in different environments: field galaxies,
binary galaxies (as identified by Peterson, 1979) and galaxy members of Zwicky
clusters (Zwicky et al., 1961-1963), subdivided into open, medium-compact and
compact clusters. They found an overabundance of unbarred early-type spirals in
dense environments, i. e. medium-compact and compact clusters, with respect to
the population of field galaxies, which appears to be in line with the
canonical galactic morphology-galactic density relation (see, e. g., the recent
review of Whitmore, 1990). But the morphological subtype population of the
barred spirals did not appear to depend on the environment; furthermore, in the
binary sample the barred and unbarred spirals were found to have a similar
subtype population.

In order to search for possible correlations between the presence of a bar in a
galaxy and its environment, EEB examined several catalogues of galaxies
pertaining to different environments: the samples of binary galaxies
constructed by Turner (1976) and Peterson (1979), the galaxy group samples
identified  by Turner and Gott (1970) and by Geller and Hucra (1983), a field
sample taken from Turner and Gott (1976) and the whole RC2 catalog, which was
taken as a general reference sample. Taking the Hubble morphological types and
the bar types for each sample from the RC2 and the UGC catalogues, EEB noted an
excess of barred early-type (Sa, Sb) spirals in the two binary samples with
respect to the other galaxy samples; this is at variance with the findings of
the Japanese authors. EEB also reported an excess of early-type spirals in
galaxy pairs for barred spirals, and realized that intrinsically fainter and
smaller binary members tend to be more frequently barred than brighter and
larger ones. Nevertheless, the statistical significance of these interesting
claims appears to be rather low (in general it is between 1 and 2 sigma).

The availability of numerous new galaxy type classifications, compiled in the
new RC3 catalogue (de Vaucouleurs et al., 1991), the fairly low statistical
significance of the above-mentioned main statistical results, and above all the
fairly vague characterization of the various galaxy samples compared (namely
the absence of a well-defined assessment of the environmental density) have
prompted us to rediscuss this matter. In \S 2 we present the galaxy sample that
we have chosen as our galaxy reference sample, i. e. the NBG catalogue (Tully
1988a). In \S 3 we discuss some suitable parameters which can be taken as
indicators of the local density of galaxies for each galaxy of the sample
considered. In \S 4 we compare the distribution functions of the various
parameters of local density for subsets of spirals of different morphological
and bar types. We also explore the environmental effects on the presence of
inner rings (which are often associated with bars) in spirals. In \S 5 we
summarize our main results. They indicate that the low-luminosity spirals
characterized by high local density tend to be both barred and early-type, thus
providing a meaningful, statistically strong confirmation of EEB's suggestions.
Taking into account the relevant N-body simulations of the effects of
interactions on spiral discs, we also discuss two alternative scenarios each of
which can account for the observational facts. On the other side, no
significant density segregation effect is observed between pure S-shaped (S(s))
and inner-ringed (S(r)) spirals.

\section{The galaxy sample}

We have chosen the 2367 galaxies of the NBG catalogue (Tully 1988a) as a galaxy
sample, since this catalogue is intended to include essentially all the nearby
galaxies with systemic velocities of less than 3000 km/sec. This corresponds to
a distance of 40 Mpc for the Hubble constant $H_0=75$ km sec$^{-1}$ Mpc$^{-1}$,
which value is adopted throughout the present paper. The potential
inhomogeneity of this catalogue is severely limited by the fact that it is
substantially a combination of two sources: (i) the magnitude-limited
Shapley-Ames sample of bright galaxies (see the RSA catalogue of Sandage e
Tammann, 1981), which assures completion across the sky to the corrected blue
total magnitude $B_T\sim 12.0$; (ii) the diameter limited sample of late-type
and fainter galaxies found in an all-sky HI survey, made principally by Fisher
and Tully (1981) and Reif et al. (1982); incompleteness becomes severe beyond
$v\sim 2000$ km/sec for this survey, which is also insensitive to the
HI-deficient objects (early-type galaxies).

In the NBG catalogue the distances of all non-cluster galaxies have been
essentially estimated on the basis of velocities, an assumed $H_0=75$ km
sec$^{-1}$ Mpc$^{-1}$, and the Virgocentric retardation model described by
Tully and Shaya (1984), in which the authors assume that the Milky Way is
retarded by 300 km/sec from the universal Hubble flow by the mass of the Virgo
cluster. The galaxy members of clusters have been given a distance consistent
with the mean velocity of the cluster (Tully 1988a).

We have updated the Hubble morphological types and the bar types (SB=barred,
SA=unbarred, SAB=transition type) of the NBG spiral galaxies by consulting the
RC3 catalogue of galaxies. In order to avoid the severe incompletion of the NBG
catalogue at large distances, in the present paper we restrict ourselves to
considering only the NBG galaxies with distance $D\leq 30$ Mpc.

\section{The local density of galaxies}

\subsection{Definition of local density parameters}

We propose to define suitable parameters which adequately express the local
density of galaxies for each spiral galaxy of our sample. Reasonably, the
incompletion of the NBG catalogue at large distances enters into the evaluation
of the local density. The incompletion of the NBG catalogue within various
shells of depth equal to 0.5 magnitude was evaluated by Tully (1988b); he
compared the observed number of galaxies with the number expected from the
normalized suitable Schechter-type (Schechter, 1976) galaxy blue luminosity
functions, using a cutoff of $M_B=-16.0$ for the galaxy blue absolute
magnitude. More than 90\% of the NBG galaxies appear to have absolute
luminosities greater than that limit. The smooth curve which describes the
observed increase in incompletion with distance $D$ (in Mpc) obeys the
expression (Tully, 1988b):
\begin{equation}
F=\exp[0.041(\mu-28.5)^{2.78}]\,,
\end{equation}
where $\mu=5\log D+25$ is the distance modulus and $F=1$ if $\mu<28.5$. The
incompletion factor $F$ expresses the number of galaxies that should have been
catalogued for each object that is listed in the NBG catalogue at a given
distance.

Tully (1988b) estimated the contribution of each galaxy brighter than $M_B=-16$
to the local density at the specific location by using a gaussian smoothing
function:
\begin{equation}
\rho_i=C\exp[-r_i^2/2(F^{1/3}\sigma)^2]\,,
\end{equation}
where $r_i$ is the spatial distance of galaxy $i$ from the specified location
and the normalization coefficient is $C=1/(2\pi)^{3/2}\sigma^3= 0.0635/
\sigma^3$. For a galaxy at a larger distance $D$, the smoothing scale length
$(F^{1/3}\sigma)$ is increased in such a way that the amplitude of the density
peak associated with a galaxy is the same at all distances $D$; in such a way
the function $\rho_i$ satisfies the normalization condition $\int\rho_idV=F$.

Each galaxy of sufficient luminosity $(M_B\leq-16)$ in the sample will
contribute to the local density at a given location in the manner described by
Eq. (2). The local density of galaxies $\rho$ at the specified location was
defined by Tully (1988b) as a summation over the contributions of all galaxies
brighter than $M_B=-16$:
\begin{equation}
\rho=\sum_i\rho_i\,.
\end{equation}
Tully (1988b) chose a smoothing constant of \sg=1 Mpc for the values of $\rho$
tabulated in his NBG catalogue. So an isolated galaxy (with $M_B\leq-16$) had a
local density of $\rho\simeq0.06$ galaxies Mpc$^{-3}$ just because of its
presence, whereas at the other extreme, in the core of Virgo cluster, the local
density $\rho$ reached the value of $\sim$~4 galaxies Mpc$^{-3}$. A smoothing
constant of \sg=0.5 Mpc was chosen for plotting the contours of surface density
of galaxies on the maps of the NBG Atlas (Tully and Fisher, 1987) (\sg=0.25 Mpc
was used for the nearest galaxies).

The above-mentioned definition of $\rho_i$ implies a fairly small dynamical
range for the local density $\rho$ of the NBG galaxies, especially when
\sg$\sim$~1 Mpc. It could be more convenient and more general to try to use a
smoothing function which diverges for $r_i\rightarrow 0$, in order to enhance
the local density $\rho$ of the galaxies which have close companions. To obtain
this we replace Eq. (2) with the following more general expression:
\begin{equation}
\rho^i_{\alpha,\sigma}=\frac{C(\alpha)}{\sigma^{3-\alpha}r_i^\alpha}
\exp[-r_i^2/2(F^{1/(3-\alpha)}\sigma)^2]\,,  \end{equation}   where $\alpha$=0,
1, 2 and the corresponding normalization coefficients are:
$C(0)=1/(2\pi)^{3/2}\simeq 0.0635$, $C(1)=1/4\pi\simeq 0.0796$, $C(2)=1/4\pi
\sqrt{2\pi}\simeq 0.0318$. In the case of a very small $r_i$, we always adopt a
lower limit of $r_i$ roughly corresponding to a typical galaxy size ($r_i$=0.05
Mpc). For $\alpha$=0 Eq. (4) simply reduces to (2).

Then we can define the galaxy local density $\rho_{\alpha,\sigma}$ of a given
NBG galaxy as the summation over all contributions of all other galaxies
brighter than $M_B=-16$, excluding the galaxy itself: \begin{equation}
\rho_{\alpha,\sigma}=\sum_i\rho^i_{\alpha,\sigma}\,. \end{equation} Thus,
according to our definitions of  $\rho^i_{\alpha,\sigma}$ and
$\rho_{\alpha,\sigma}$ (Eqs. (4) and (5)), an isolated galaxy always has
$\rho_{\alpha,\sigma}$=0, for any $\alpha$ and \sg, whilst, according to the
previous expressions (2) and (3), it would have a local density $\rho$
unreasonably dependent on the choice of \sg.

In general, the choice of low $\sigma$-values lead to large $\rho$-values,
because of the clustering properties of galaxies, as we shall clarify in the
next subsection. We have verified that, compared to $\rho_{0,\sigma}$, the
parameters $\rho_{1,\sigma}$ and $\rho_{2,\sigma}$ remain much more sensitive
to the presence of very close neighbours also for fairly large $\sigma$-values.
Thus, even for fairly high $\sigma$-values, $\rho_{1,\sigma}$ and
$\rho_{2,\sigma}$ behave in a manner substantially similar to $\rho_{0,\sigma}$
with fairly low $\sigma$-values. Therefore, in the following we shall limit
ourselves to considering \ras in the case $\alpha$=0, denoting this quantity
simply by $\rho_\sigma$.

In order to calculate the local density \rs we need to know the absolute
magnitudes of the NBG galaxies, of which those with $M_B\leq-16$ will be taken
as contributors to the local density. We used the $M_B$ estimates of many NBG
galaxies (as recorded in the NBG) based on the adopted distances and the
corrected total blue apparent magnitudes. We estimated the absolute magnitudes
of several galaxies with unknown apparent magnitudes from their corrected
isophotal diameters $D_{25}$ (relative to the 25 $B_T$ mag  arcsec$^{-2}$
brightness level) tabulated in the NBG catalogue, by relying on the following
standard luminosity-diameter relations: \begin{equation} M_B=-4.8\log
D_{25}-13.8\,, \end{equation} with $D_{25}$ expressed in kpc, for the
elliptical galaxies (Giuricin et al., 1989), and \begin{equation} M_B=-5.7\log
D_{25}-12.4 \end{equation} for the lenticular and spiral galaxies (Girardi et
al., 1991).

Another suitable parameter of the local density of galaxies that we shall use
in this paper is the number of galaxies, always with $M_B\leq-16$, which are
located within a spherical volume of radius $RF^{1/3}$ around a given galaxy,
divided by the volume $4\pi R^3/3$; $F$ again takes into account the
incompleteness of the sample. We denote the resulting local galaxy density as
$C_R$ (expressed in galaxies Mpc$^{-3}$).

Other useful indices of the local galaxy density are the distances (expressed
in Mpc) $d_1$ ($d_2$, $d_3$\ldots) of the first (second, third\ldots) nearest
galaxy, the mean distances $\langle d_n\rangle$ of the first $n$ nearest
galaxies; another interesting quantity is the morphological type $T_n$ of the
$n$th nearest galaxy.

\subsection{The Behaviour of \rs as a function of \sg}

As already mentioned, the choice of low \sg values leads to large $\rho$
values, because of the clustering properties of galaxies, as we shall clarify
now. Notably, the mean value of the parameter \rs, calculated for the galaxies
of our sample, is related to the two-point correlation function of galaxies,
$\xi(r)$, defined in terms of the probability $dP$ of finding that a galaxy,
chosen arbitrarily from our galaxy sample, has a neighbour in the infinitesimal
volume $dV$ at distance $r$: \begin{equation} dP=n(1+\xi(r))dV\,,
\end{equation} where $n=N/V$ is the mean number of galaxies found within the
finite volume $V$. As a matter of fact, the contribution to the local density
\rs of the volume $dV$ will be $dP\cdot\rho^i_\sigma$, with  $\rho^i_\sigma$
given by Eq. (4). The average local density $\langle\rho_\sigma\rangle$ for the
galaxies of our sample will be \begin{equation}
\langle\rho_\sigma\rangle=\int_V(1+\xi(r))\rho_\sigma^i({\bf r})dV\,.
\end{equation} The integral is extended over the volume occupied by our galaxy
sample, which in our case, because of the zone obscured by the Milky Way plane,
is roughly \begin{equation} V=\frac{4}{3}\pi D^3(1-\sin\theta)\simeq 8.8\cdot
10^4\ Mpc^3\,, \end{equation} with $D$=30 Mpc and $\theta\sim15^\circ$;
2$\theta$ is the adopted mean angular extension of the obscuration zone.

The above-mentioned quantity $n=N/V\simeq 0.019$ galaxies Mpc$^{-3}$ is the
mean density of galaxies, uncorrected for the incompleteness of the galaxy
sample. The corrected mean density of galaxies is $n'=N'/V\simeq 0.50$ galaxies
Mpc$^{-3}$, with $N'=\sum_{i=1}^{N}F_i=4165.4$, where the incompletion factor
$F_i$ for the $i$th galaxy is given by Eq. (1). We now define the mean value
$\langle F\rangle=\sum_{i=1}^{N}F_i/N=N'/N$; replacing the function $F$
contained in the Eq. (4) with its mean value $\langle F\rangle$, and using the
normalization condition $\int\rho_\sigma^idV=\langle F\rangle$, the quantity
\begin{equation}
\Delta\rho_\sigma=(\langle\rho_\sigma\rangle-n')/n'
\end{equation}
can be written as:
\begin{equation}
\Delta\rho_\sigma\simeq\frac{1}{\langle F\rangle}\int_V\xi(r)
\rho^i_\sigma({\bf r})dV\,.
\end{equation}

For a canonical power-law correlation function $\xi(r)=(r/r_0)^{-\gamma}$ we
obtain the following analytical expression for $\Delta\rho_\sigma$:
\begin{equation}
\Delta\rho_\sigma\simeq(4\pi r_0^\gamma/((2\pi)^{3/2}\sigma^3\langle
F\rangle))\int_0^\infty r^{2-\gamma}\exp[-r^2/2(\langle F\rangle^{1/3}
\sigma)^2]dr\,.
\end{equation}
$\Delta\rho_\sigma$ can be rewritten in terms of the new variable
$x=r/\sqrt{2}\langle F\rangle^{1/3}\sigma$ as
\begin{equation}
\Delta\rho_\sigma\simeq g(\gamma)(\sigma/r_0)^{-\gamma}
\end{equation}
with
\begin{equation}
g(\gamma)=\frac{4\pi}{(2\pi)^{3/2}}2^{(3-\gamma)/2}\langle F\rangle^{-\gamma}
\int_0^\infty x^{2-\gamma}\exp{-x^2}dx\,.
\end{equation}

Eq. (14) clearly shows that, for the reasonable case of $\gamma>0$, low \sg
values give large values of  $\Delta\rho_\sigma$ (and $\langle\rho_\sigma
\rangle$), as anticipated above. Furthermore, we expect a roughly linear
$\log\Delta\rho_\sigma-\log\sigma$ relation, with a slope equal to $\gamma$.
Table 1 presents the values, for our galaxy sample, of the average local
densities $\langle\rho_\sigma\rangle$ (in galaxies Mpc$^{-3}$) and  $\Delta
\rho_\sigma$ relative to different \sg values. Fig. 1 illustrates the
expected approximate linear relation between  $\log\Delta\rho_\sigma$ and
$\log\sigma$, with $\gamma\sim 1$ and the correlation length $r_0\sim 19$ Mpc.
There seems to be a discrepancy between the values of $\gamma$ and $r_0$ which
fit the points plotted in Fig. 1 and the canonical values $\gamma\sim 1.8$ and
$r_0$ in the range 5$\div$11 Mpc (for our $H_0$ of 75 km sec$^{-1}$
Mpc$^{-1}$), which describe the two-point correlation function $\xi(r)$ of
individual galaxies (see, e. g., the review of Geller, 1987). This discrepancy
is not unexpected, since our sample of nearby galaxies is not a fair sample of
the universe.

\subsection{Correlations between different parameters of local density}

{}From our galaxy sample we have analyzed many correlations between the
different
parameters of local density defined above, in order to clarify their meaning.
Fig. 2 displays the correlations between $\rho_{1.0}$ (i. e. \sg=1 Mpc) and
$\rho_{0.5}$, $\rho_{1.25}$ and $\rho_{1.5}$. The larger the difference between
the \sg values, the poorer the correlation obviously is, but it is remarkable
that the points tend to be located on different main relations; this is due to
the presence of some clusters of galaxies, like the Virgo Cluster, whose
members (denoted by squares in Fig. 2) tend to stay mainly on the relations
which go up to the largest $\rho_{1.0}$ values. In other words, the members of
clusters tend to be characterized by large $\rho$ values whenever large \sg
values (\sg$\sim$ 1 Mpc) are adopted, whereas binary galaxies or galaxies with
very close neighbours, at distances of a few tenths of Mpc, will have high
$\rho$ values particularly if low \sg values (\sg$\sim$ 0.25 Mpc) are adopted.

Examining the correlations between $\rho_{0.5}$ and the densities of galaxy
counts $C_R$ for various $R$-values we have found that the best correlation is
with $C_{1.0}$ (see Fig. 3). We have verified that in general \rs tends to show
the best correlation with $C_R$ when $R\sim 2\sigma$, and that these
correlations appear to be rather poor for low $R$ values, because of the
discreteness of the galaxy counts, and tend to become better and better for
larger $R$. As expected, there is a good correlation between $\rho_{0.1}$ and
$d_1$ (the distance of the nearest galaxy), whilst the correlations between the
same density parameter and $d_2$  becomes poorer (see Fig. 4). We have also
found that the distances and mean distances of the somewhat farther companions
better correlate with the values of \rs relative to the fairly
large \sg values.

\section{Results}

\subsection{Barred spirals}

Table 2 lists the number of spiral galaxies found, binned by Hubble
morphological type (Sa--Sab, Sb--Sbc, Sc--Scd and Sd--Sm) and bar type (SA, SAB
and SB),  for the 866 NBG spirals with $M_B\leq-16$, distance $D\leq30$ Mpc and
known bar classification, according to the RC3 catalogue. We have calculated
the distribution functions of the parameters \rs for \sg=0.25, 0.5, 1.0 and 2.0
Mpc, for the above-mentioned subsets of galaxies with different Hubble types
and bar classifications. Figs. 5, 6, 7 and 8 show the cumulative distribution
functions of \rs for \sg=0.25 and 1.0 Mpc. Clearly, early-type barred spirals
(SBa and SBab) tend to have greater local densities \rs than unbarred ones (SAa
and SAab), especially for \sg=0.25 Mpc; moreover, as the distribution function
of SAB galaxies is more similar to that of SBs, the reasonable inclusion of the
SAB galaxies into the SB subset will strengthen this difference. On the other
hand, no appreciable density segregation is observed in the later morphological
types.

In order to test the significance of the differences between the distributions
of \rs for the two different subsets of galaxies, the SAa-ab and SBa-ab plus
SABa-ab subsets,we have applied the classical Kolmogorov-Smirnov test
(hereafter KS; e. g. Hoel, 1971), the Rank-Sum test (hereafter RS; e. g. Hoel,
1971), and the Mann-Whitney U-test (e. g. Kendall and Stuart, 1979). In Table 3
we give the probabilities in percent, p(KS), p(U) and p(RS), that the two
distributions of \rs, for the two subsets defined above, refer to different
distribution functions according to the three (two-tailed) statistical tests.
Table 3 specifies that the above-mentioned density effect is particularly
strong (at the $\sim$3 sigma significance level) for \rs with low \sg values
(mainly  \sg=0.25) and tends to weaken as we move to higher \sg values. This
indicates that this effect is mainly due to the presence of close companions,
at a distance of a few tenths of Mpc. We have verified that no appreciable
density segregation between barred and unbarred objects is observed in the mid-
and late-type spirals. Furthermore, the 32 S0/a galaxies have not been included
in the Sa--Sab subset because they do not show any segregation and are too few
to be considered as a subset by themselves.

The use of the density parameter $C_R$, for $R$=0.2, 0.5, 1.0 and 2.0 Mpc,
leads to similar results: the barred early-type (Sa, Sab) spirals appear to
stay  appreciably in zones of higher local density, mainly for $R$=0.5 Mpc,
compared to the unbarred early-type ones (see Table 4). If in the calculation
of $C_R$ we restrict ourselves to counting separately the E/S0, early-type
(S0/a--Sb) spirals and late-type (Sbc--Sm) spirals, we realize that the density
effect is substantially due to the presence of early-type galaxy neighbours,
E/S0 and early-type spirals. This finding is likely to be related to the
well-known preference of early-type galaxies to occupy denser regions.

The evaluation of the distances $d_1$, $d_2$, $\langle d_2\rangle$ and
$\langle d_3\rangle$ confirms the previous results: as a matter of fact, within
the early-type spirals (Sa--Sab) the barred spirals have, on average, smaller
distances from the nearest neighbours than the unbarred ones (see Fig. 9 and
Table 5). Furthermore, from the distributions of the morphology $T_1$ of the
nearest neighbour, we infer that the barred early-type spirals tend to have
galaxies of an earlier type as companions, if compared to the unbarred ones
(see Fig. 10 and Table 5); this is consistent with our previous findings.

When we divide our early-type spirals into objects intrinsically brighter or
fainter than $M_B$=--20, we find that the local density segregation between
barred and unbarred spirals essentially concerns only the less luminous objects
(see Fig. 11 and Table 6), whereas no appreciable effect is apparent in the
high-luminosity range.

After having divided the SA, SAB and SB spirals into three intervals of local
density, $\rho_{0.25}<0.001$, $0.001\leq\rho_{0.25}\leq 3$, $\rho_{0.025}>3$
galaxies Mpc$^{-3}$, we have examined whether the morphological type
distributions of each of the three bar types are different in the three density
bins. The arbitrary choice of the three density intervals is related to the
convenience of having a comparable number of objects in the three density
intervals; nevertheless, we have verified that lowering the upper limit of 3
galaxies Mpc$^{-3}$ to 2 does not significantly change the results. We have
found that the morphological type distribution of the SB galaxies associated
with the highest density range is significantly shifted, at a level of $\sim$3
sigma, to morphological types earlier than the SB located in the lower density
regions (see Fig. 12 and Table 7), whereas no morphological type shift as a
function of the local density is observed for the SA and SAB types. Finally, we
have noted a similar behaviour using other \rs parameters; for instance, for
$\rho_{1.0}$, taking as bin limits 0.1 and 0.3 galaxies Mpc$^{-3}$, the SB
galaxies in the densest zones show a $\sim$3 sigma excess of early type
spirals, whereas for the SA and SAB galaxies the distributions of morphological
type are substantially constant for the various density regions.

\subsection{Ringed spirals}

The occurrence of rings in galaxies, generally distinguished into nuclear,
inner and outer rings, is observed to be frequently associated with the
presence of bars (see, e. g., the review of Buta, 1989, on the observational
properties of rings). Here we shall consider only the inner rings, which are
more widely present and more easily discernible than the other two types of
rings; in this case barred and unbarred galaxies are taken together. The
presence of inner rings defines the so-called S(r) variety of the standard
galaxy classification adopted in the RC2 and RC3 catalogues. Various dynamical
mechanisms have been advanced to explain the formation of inner (and nuclear as
well) rings: viscous torques, outflowing winds thermally driven by violent star
formation rates, gravitational torques (and Lindblad resonances) requiring
non-axisymmetric potentials which are thought to be mostly due to bars or tidal
interactions (see, e. g., the review of Combes, 1991, on the theoretical
aspects of ring formation). Lastly, Zaspel (1992) stressed that ring structures
can arise in a gas or stellar disk without assuming the initial presence of a
bar or oval distorsion. This would explain the existence of some disk galaxies
with definite ring structures, but very weak (if any) bar features (as observed
by Buta, 1990a, b).

If ring-like features are often triggered during a tidal interaction, as is
shown in some N-body simulations (Combes, Dupraz and Gerin, 1990; Combes,
1991), we expect to find an overabundance of ringed systems (S(r) spirals)
located in high-density regions with respect to the pure spiral S-shaped (S(s))
systems.

Table 8 contains the number of NBG spirals within a distance of 30 Mpc, binned
by Hubble morphological type and ring type (S(r), S(rs), S(s)). Comparing the
distribution functions of the parameters \rs for \sg=0.25, 0.5, 1.0 and 2.0 Mpc
for the various subsets of galaxies, we have found no significant density
segregation between S(r) (or S(r)+S(rs)) and S(s) spirals, either for the whole
spiral sequence or for the early types alone (see Fig. 13; we have checked that
the weak difference between the distributions shown in this figure is not
statistically significant ($<$1 sigma)). The subdivision of our S(r), S(rs),
S(s) galaxies into high-luminosity and low-luminosity objects (with $M_B=-20$
as a limit) always confirms this negative result. We conclude that tidal
interactions probably play a minor role (if any) in the formation of inner
rings, compared to the other mechanisms mentioned above. Nevertheless, we are
aware that the categorization of spirals into the S(r), S(rs) and S(s)
varieties is less secure than the bar type classification and that independent
observers are often not very consistent about this subtle aspect of galaxy
structure.

\section{Discussions and conclusions}

Employing a variety of parameters for local galaxy density, we have found a
significant excess of barred galaxies in the high-density regions (on scales of
a few tenths of Mpc) of our nearby universe, for the early-type (Sa, Sab) and
relatively faint ($M_B\geq-20$) spirals, together with an overabundance of
early-type spirals in the subset of the barred systems which reside in denser
regions. Thus, spiral galaxies which inhabit dense regions, on scales of
$\sim$0.25 Mpc, tend to be barred if they are of early type and tend to be of
early type if they are barred. Consequently, the extension of the canonical
morphology--density relation to the spiral sequence --- according to which
early-type spirals tend to reside in denser regions than late-type objects (e.
g. Giuricin et al., 1988; Tully, 1988b) --- appears to hold for the barred
family alone. Our findings provide a meaningful, statistically strong
confirmation of the recent, similar (albeit weaker) results of EEB, which,
however, lack good criteria for the definition of the local density of
galaxies.

The fact that the density segregation that we have found is seen essentially in
low-luminosity spirals can be easily explained as follows. As a matter of fact,
several N-body simulations concur that interactions stimulate or considerably
accelerate the formation of a bar in a galaxy, particularly in the case of a
more massive companion (Noguchi, 1987; Gerin et al., 1990; Salo, 1991) --- i.
e., especially for faint galaxies, on average.  Besides, interaction-induced
bar formation is theoretically expected to be easier in galaxy models which
have rising rotation curves up to a large fraction of the galaxy optical disk
radius --- as in low-luminosity spirals (see e. g. Persic and Salucci, 1991)
--- than in models which have rotation curves which rise up to a small fraction
of the disk and become flat thereafter --- as in high-luminosity spirals. In
the latter galaxy models the preferred outcome of an interaction appears to be
an increase in the rate of disruptive collisions between high-velocity
interstellar gas clouds, with consequent enhanced star formation, especially in
the central galactic region, without a formation of strong, long-lived bars
(Olson and Kwan, 1990).

Nevertheless, on the basis of the available numerical simulations it is not
easy to understand why galaxy interactions promote bar formation primarly in
early-type spirals. We speculate that this different behaviour of early and
late-type spirals can be related to the fact that the bars in early types have
basically different photometric and kinematic properties (being generally
stronger and longer) than those in late types, probably ending near different
orbital resonances (see, e. g., Elmegreen and Elmegreen, 1985, 1989).

In order to account for this different behavioure, EEB appear to prefer a
scenario in which close encounters of galaxies are supposed to induce both bar
formation in a galaxy and a simultaneous shift of its morphological type
towards early types. The latter effect could be caused by interaction-induced
inflow of stellar and gaseous mass --- which is found in the N-body simulations
of Byrd et al. (1986) and Byrd, Sundelius, Valtonen (1987). This inflow would
increase the stellar and gas density in the inner regions; besides, in the case
of prolonged interactions, the inflow of the gaseous dissipative component
would deplete the outer disk of gas and lower the star formation rate in the
outer regions. These changes would correspond to an alteration of the galaxy
morphology towards earlier types. Within this scenario the morphology--density
relation in the spiral sequence would be mostly due to interaction-induced
evolutionary processes, rather than to density-dependent processes of galaxy
formation. The recent N-body simulations of Noguchi (1988) and Salo (1991)
confirm that tidally induced stellar bar structures are able to drive an
efficient infall of the gas component in the disk to the nuclear region; this
infall would trigger nuclear activity in interacting galaxies (see, e. g., the
review of Heckman, 1990). But they also find no appreciable corresponding
large-scale infall of the non-dissipative stellar component. However, a gas
infall would give rise to star formation, and if this process is sufficiently
fast and efficient it could lead to an increase of the stellar component of the
bulge; as a consequence, the bulges of barred early types would tend to appear
somewhat bluer than the unbarred ones, a fact which has not been observed yet.
In conclusion, interaction-induced changes in the galactic morphological type
do not appear to be well documented by the most recent N-body simulations.

Alternatively, our observational results could be framed in a view which does
not involve any interaction-induced alteration of the galactic morphology.
Unfortunately, the basic galaxy structural parameters which are generally
varied in the available simulations (such as the disc-to-halo mass ratio and
the degree of central concentration of the halo mass) do not describe properly
spiral galaxies of different morphological types; therefore, the effect of
interactions on bar triggering as a function of morphological type can not be
inferred. Nevertheless, we could envisage a scenario in which interactions are
able to trigger prominent bar structures essentially in early types; besides,
the excess of early types in barred systems associated with high local
densities could be explained if we admitted that the galaxies in denser regions
are preferentially early-type because of a morphology-density relation:
eventually this excess in unbarred spirals would be cancelled by the fact that
interacting early-type SAs have become barred at later times. On this view it
is implicit that the morphology-density relation is likely to be due to the
initial conditions for galaxy formation --- see, e. g., the ``biased'' scheme
of galaxy formation (Blumenthal et al., 1984) --- or to global (and local)
processes occurring at the beginning of galaxy life --- see, e. g., the
density-dependent rate of conversion of gas into stars after galaxy collapse
(Icke, 1985; Lacey and Silk, 1991) ---  rather than to late evolutionary
processes.

The two basic scenarios that we have outlined require dramatic (possibly
unrealistic) processes, i. e. a radical change in morphology and a drastic
dependence of bar stimulation on the morphology itself. A scenario, in which
both processes occur simultaneously, albeit in a weaker way, seems more
reasonable: bar triggering, easier for an early-type spiral, accompanied by a
slight change in morphology, could lead to the almost discontinuous behaviour
observed. In this case a primordial morphology-density relation would still be
needed.

\bigskip
\bigskip

The authors thank Massimo Persic and Paolo Salucci for useful discussions. One
of the authors (P. M.) acknowledges enlightening conversations with
Fran\c{c}oise Combes. The authors are also  grateful to Harold G. Corwin Jr.
for having provided them with the ninth tape version of the Third Reference
Catalogue of Bright Galaxies (RC3). This work was partially supported by the
Ministry of University and Scientific and Technological Research (MURST) and by
the Italian Research Council (CNR-GNA).

\newpage

\section*{Tables}

{\bf Table 1:} The quantities $\langle\rho_\sigma\rangle$ and
$\Delta\rho_\sigma$ for different \sg values.

\bigskip

\begin{center}
\begin{tabular}{|l|c|c|c|} \hline
$\sigma$ & $\langle\rho_\sigma\rangle$ & $\Delta \rho_\sigma$\\
(Mpc) & (gal/Mpc$^3$) & --- \\ \hline
0.1  & 2.46  & 48.53\\
0.25 & 1.43  & 27.85\\
0.5  & 0.79  & 14.87\\
0.75 & 0.52  & 9.52\\
1.0  & 0.38  &\ 6.58\\
1.25 & 0.28  &\ 4.73\\
1.5  & 0.22  &\ 3.49\\
2.0  & 0.15  &\ 2.01\\ \hline
\end{tabular}
\end{center}

Table 1: Values of the mean density $\langle\rho_\sigma\rangle$ (in
galaxies Mpc$^{-3}$) and of the quantity $\Delta\rho_\sigma$, as defined in
Eq.  (11), calculated for the NBG galaxies with $M_B\leq-16$ and $D\leq30$ Mpc,
for the indicated values of \sg (in Mpc).

\newpage

{\bf Table 2:} The galaxy sample: barred and unbarred galaxies.

\bigskip

\begin{center}
\begin{tabular}{|c|llll|} \hline
    & Sa, Sab &  Sb, Sbc & Sc, Scd & Sd--Sm\\ \hline
\hline
SA  &  34(36) & 66(34) & 80(32) & \ 33(10)\\
SAB &  25(26) & 65(34) & 83(33) & \ 83(25)\\
SB  &  36(38) & 61(32) & 90(35) & 210(65) \\ \hline
TOT &  95     & 192    & 253    & 326 \\ \hline
\end{tabular}
\end{center}

Table 2: Number of NBG galaxies with $M_B\leq-16$, $D\leq30$ Mpc and known bar
type according to RC3, divided into four morphological type bins. The relative
percentages are given in parenthesis.

\bigskip
\bigskip

{\bf Table 3:} Comparison of SA and SAB, SB for early-type spirals: \rs.

\bigskip

\begin{center}
\begin{tabular}{|c|c|cccc|} \hline
\sg=   &       &  0.25 &  0.5  &  1.0  &  2.0 \\ \hline
SA     & p(KS) & 99.89 & 97.18 & 98.22 & 83.04\\
vs     & p(U)  & 99.98 & 99.29 & 95.73 & 92.77\\
SB+SAB & p(RS) & 99.94 & 99.26 & 95.72 & 92.76\\ \hline
\end{tabular}
\end{center}

Table 3: Results of the KS, U and RS tests for the comparison of the
distributions of local density $\rho_\sigma$ of the two indicated subsets of
Sa, Sab galaxies, for the indicated values of \sg. The three numbers p(KS),
p(U) and p(RS) give the probability, in percent, that the two distributions
compared are different.

\newpage

{\bf Table 4:} Comparison of SA and SAB, SB for early-type spirals: $C_R$.

\bigskip

\begin{center}
\begin{tabular}{|c|c|cccc|} \hline
$R$=   &       &  0.2  &  0.5  &  1.0  &  2.0 \\ \hline
SA     & p(KS) & 29.73 & 99.65 & 78.78 & 76.25\\
vs     & p(U)  & 96.72 & 99.97 & 98.63 & 93.57\\
SB+SAB & p(RS) & 78.14 & 99.88 & 98.52 & 93.54\\ \hline
\end{tabular}
\end{center}

Table 4: Results of the KS, U and RS tests for the comparison of the
distributions of local density $C_R$ of the two indicated subsets of Sa, Sab
galaxies, for the indicated values of $R$. The three numbers p(KS), p(U) and
p(RS) give the probability, in percent, that the two distributions compared are
different.

\newpage

{\bf Table 5:} Comparison of SA and SAB, SB for early-type spirals:
distances.

\bigskip

\begin{center}
\begin{tabular}{|c|c|ccccc|} \hline
&&$d_1$ & $d_2$ & $\langle d_2\rangle$ & $\langle d_3\rangle$ & $T_1$\\ \hline
SA     & p(KS) & 99.98 & 92.78 & 99.33 & 95.19 & 79.18\\
vs     & p(U)  & 99.94 & 98.80 & 99.65 & 99.38 & 97.97\\
SB+SAB & p(RS) & 99.94 & 98.80 & 99.64 & 99.38 & 97.88\\ \hline
\end{tabular}
\end{center}

Table 5: Results of the KS, U and RS tests for the comparison of the
distributions of local distances $d_1$, $d_2$, $\langle d_2\rangle$, $\langle
d_3\rangle$ and $T_1$ (morphological type of the nearest companion) of the two
indicated subsets of Sa, Sab galaxies. The three numbers p(KS), p(U) and p(RS)
give the probability, in percent, that the two distributions compared are
different.

\newpage

{\bf Table 6:} Comparison of SA and SAB, SB for early-type faint spirals:
\rs.

\bigskip

\begin{center}
\begin{tabular}{|c|c|cccc|} \hline
\sg=   &       &  0.25 &  0.5  &  1.0  &  2.0 \\ \hline
SA     & p(KS) & 99.95 & 99.51 & 95.34 & 72.23\\
vs     & p(U)  & 99.98 & 99.85 & 98.05 & 90.34\\
SB+SAB & p(RS) & 99.96 & 99.84 & 98.06 & 90.34\\ \hline
\end{tabular}
\end{center}

Table 6: Results of the KS, U and RS tests for the comparison of the
distributions of local density $\rho_\sigma$ of the two indicated subsets of
Sa, Sab galaxies fainter than $M_B=-20$, for the indicated values of \sg. The
three numbers p(KS), p(U) and p(RS) give the probability, in percent, that the
two distributions compared are different.

\newpage

{\bf Table 7:} Comparison of the morphological type distributions for
spirals in different density regions.

\bigskip

\begin{center}
\begin{tabular}{|c|c|rrr|} \hline
    &       & ld vs md & md vs hd & ld vs hd\\ \hline \hline
    & p(KS) &  \ 3.63  &  \ 0.04  &   30.65 \\
SA  & p(U)  &   66.35  &   33.79  &   28.62 \\
    & p(RS) &   65.74  &   33.34  &   28.24 \\ \hline \hline
    & p(KS) &   29.02  &  \ 0.03  &   50.76 \\
SAB & p(U)  &   60.08  &  \ 9.77  &   66.75 \\
    & p(RS) &   59.68  &  \ 9.70  &   66.30 \\ \hline \hline
    & p(KS) &   47.08  &   98.70  &   97.95 \\
SB  & p(U)  &   65.06  &   99.53  &   99.54 \\
    & p(RS) &   64.50  &   99.48  &   99.50 \\ \hline
\end{tabular}
\end{center}

Table 7: Results of the KS, U and RS tests for the comparison of the
distributions of morphological types T, coded as in NBG, for the three subsets
of low density (ld, $\rho_{0.25}<0.001$), mid density (md,
$0.001\leq\rho_{0.25}\leq 3$) and high density (hd, $\rho_{0.25}>3$) galaxies,
divided in SA, SAB and SB. The three numbers p(KS), p(U) and p(RS) give the
probability, in percent, that the two distributions compared are different.

\newpage

{\bf Table 8:} The galaxy sample: inner-ringed (S(r)) and unringed (S(s))
galaxies.

\bigskip

\begin{center}
\begin{tabular}{|c|llll|} \hline
    & Sa, Sab & Sb, Sbc &  Sc, Scd & Sd--Sm\\ \hline
Sr  & 24(26)  & 37(20)  & \ 13(\ 6)& \ \ 4(\ 1)\\
Srs & 20(21)  & 77(42)  & \ 90(38) & \ 43(16)\\
Ss  & 49(53)  & 70(38)  &  134(56) & 227(83) \\ \hline
TOT & 93      & 150     &  237     & 274     \\ \hline
\end{tabular}
\end{center}

Table 8: Number of NBG galaxies with $M_B\leq-16$, $R\leq30$ Mpc and known
inner ring type according to RC3, divided in four morphological type bins. The
relative percentages are given in parenthesis.

\newpage

\section*{Figure captions}

{\bf Figure 1:} Plot of $\log \Delta\rho_\sigma$ versus $\log\sigma$ for our
sample of NBG galaxies with $M_B\leq-16$ and distance $D\leq 30$ Mpc. The
straight line fitting the data points is indicated.

\bigskip

{\bf Figure 2:} Plots of (a) $\rho_{0.5}$, (b) $\rho_{1.25}$ and (c)
$\rho_{1.5}$ versus $\rho_{1.0}$. The galaxies of the Virgo Cluster are denoted
by squares, whereas the other galaxies are denoted by crosses.

\bigskip

{\bf Figure 3:} Plots of the densities of galaxy counts $C_{1.0}$ versus
$\rho_{0.5}$.

\bigskip

{\bf Figure 4:} Plots of (a) $d_1$ (distance of the nearest galaxy) and (b)
$d_2$ (distance of the second nearest galaxy) versus $\rho_{0.1}$.

\bigskip

{\bf Figure 5:} Plots of the cumulative distribution functions of the local
density parameters (a) $\rho_{0.25}$ and (b) $\rho_{1.0}$ for the Sa, Sab
spiral galaxies. The distributions for the bar types SA, SAB, SB  are indicated
by solid, dotted, dashed lines respectively.

\bigskip

{\bf Figure 6:} Plots of the cumulative distribution functions of the local
density parameters (a) $\rho_{0.25}$ and (b) $\rho_{1.0}$ for the Sb, Sbc
spiral galaxies. The distributions for the bar types SA, SAB, SB  are indicated
by solid, dotted, dashed lines respectively.

\bigskip

{\bf Figure 7:} Plots of the cumulative distribution functions of the local
density parameters (a) $\rho_{0.25}$ and (b) $\rho_{1.0}$ for the Sc, Scd
spiral galaxies. The distributions for the bar types SA, SAB, SB  are indicated
by solid, dotted, dashed lines respectively.

\bigskip

{\bf Figure 8:} Plots of the cumulative distribution functions of the local
density parameters (a) $\rho_{0.25}$ and (b) $\rho_{1.0}$ for the Sd, Sdm, Sm
spiral galaxies. The distributions for the bar types SA, SAB, SB  are indicated
by solid, dotted, dashed lines respectively.

\bigskip

{\bf Figure 9:} Plots of the cumulative distribution functions of the local
density parameters (distances, in Mpc) (a) $d_1$ and (b) $d_2$ for the Sa, Sab
spiral galaxies. The distributions for the bar types SA, SAB, SB  are indicated
by solid, dotted, dashed lines respectively.

\bigskip

{\bf Figure 10:} Plot of the cumulative distribution functions of the
parameter $T_1$ (morphological type of the nearest galaxy, coded as in NBG
catalogue) for the Sa, Sab spiral galaxies. The distributions for the bar types
SA, SAB, SB  are indicated by solid, dotted, dashed lines respectively.

\bigskip

{\bf Figure 11:} Plot of the cumulative distribution functions of the local
density parameter $\rho_{0.25}$ for the Sa, Sab spiral galaxies (a) brighter
and (b) fainter than $M_B=-20$. The distributions for the bar types SA, SAB, SB
are indicated by solid, dotted, dashed lines respectively.

\bigskip

{\bf Figure 12:} Plots of the distribution functions of the morphological type
T, coded as in NBG, for the (a) SA, (b) SAB and (c) SB spirals binned in three
different local density ranges: $\rho_{0.25}<0.001$ (ld), $0.001\leq\rho_{0.25}
\leq 3.0$ (md), $\rho_{0.025}>3$ galaxies Mpc$^{-3}$ (hd), denoted by solid,
dotted, dashed lines respectively.

\bigskip

{\bf Figure 13:} Plots of the cumulative distribution functions of the local
density parameter $\rho_{0.25}$ for the Sa, Sab spiral galaxies. The
distributions for the ring types S(r), S(rs), S(s)  are indicated by solid,
dotted, dashed lines respectively.

\end{document}